\documentclass{svproc}
\usepackage{url}

\usepackage{graphicx}
\usepackage{lineno}
\usepackage{amsmath}

\begin{document}
\mainmatter              
%
\title{Beauty Production with ALICE at the LHC}
\titlerunning{Beauty production with ALICE}  
%
\author{Erin F. Gauger for the ALICE Collaboration}
\authorrunning{E. Gauger for the ALICE Collaboration} 
%
%
\institute{The University of Texas at Austin, Austin TX 78712, USA,\\
\email{erin.gauger@cern.ch}}

\maketitle              

\begin{abstract}
In this manuscript, various beauty production measurements using the ALICE detector will be presented. We will show new measurements of non-prompt D$^0$ mesons in pp collisions at $\sqrt{s} = 5.02$ TeV and beauty-tagged jet production in p--Pb collisions at $\sqrt{s_{\mathrm{NN}}} = 5.02$ TeV. The $R_{\mathrm{AA}}$ of beauty-hadron decay electrons in central Pb--Pb collisions and the $v_{2}$ of beauty-hadron decay electrons in semi-central Pb--Pb collisions at $\sqrt{s_{\mathrm{NN}}} = 5.02$ TeV will also be discussed.
\end{abstract}
\section{Introduction}
In hadronic collisions, beauty quarks are produced early via hard-scattering processes with large momentum transfer. Because of this early production, the beauty quark is an excellent probe of the Quark-Gluon Plasma (QGP) \cite{qgp1,qgp2} formed in heavy-ion collisions. Traveling through the QGP, beauty quarks interact with other partons via collisional and radiative processes and lose energy. The beauty quark ($m_{\mathrm{b}} \simeq 4.18$ GeV/$c^2$ \cite{pdg}) is expected to lose less energy than lighter quarks, since collisional processes depend on the mass of the particle and the dead cone effect \cite{deadCone} would hamper radiative energy loss. Since charm (${m_{\mathrm{c}} \simeq 1.27\text{ GeV}/c^2}$ \cite{pdg}) and beauty quarks are both produced early in the collision but have different masses, it is useful to compare beauty and charm measurements to test our understanding of mass-dependent energy loss in heavy-ion collisions. 

Measurements of beauty production in pp collisions are important to test perturbative QCD (pQCD) calculations, as well as to provide the baseline for Pb--Pb measurements. In p--Pb collisions, beauty-production measurements are crucial to isolate initial-state and cold nuclear matter effects, both of which would be present in Pb--Pb measurements. 
\section{Beauty measurements}
In ALICE, beauty production is measured via beauty-hadron decay electrons, non-prompt D$^{0}$ mesons, and beauty-tagged jets. All three measurements rely on the excellent vertex reconstruction and impact parameter resolution of the ALICE detector to isolate the decay particles and jets from beauty decay. 
\subsection{Beauty-hadron decay electrons}
Roughly 10\% of beauty hadrons decay directly into electronic final states (e.g. B$^{-}$$\rightarrow$ e+X), and another 10\% decay to charm hadrons which further decay to electrons (e.g. B$^{-}$$\rightarrow$ D$^{0}$$\rightarrow$ e+X) \cite{pdg}. This high branching ratio coupled with the excellent electron identification of the ALICE detector makes it convenient to study beauty quarks by measuring the production of beauty-hadron decay electrons. Beauty-hadron decay electrons have been measured in pp, p--Pb, and Pb--Pb collisions \cite{bepp,bepp276,bePbPb}.

\begin{figure}[h!]
\begin{center}
\includegraphics[width=6cm]{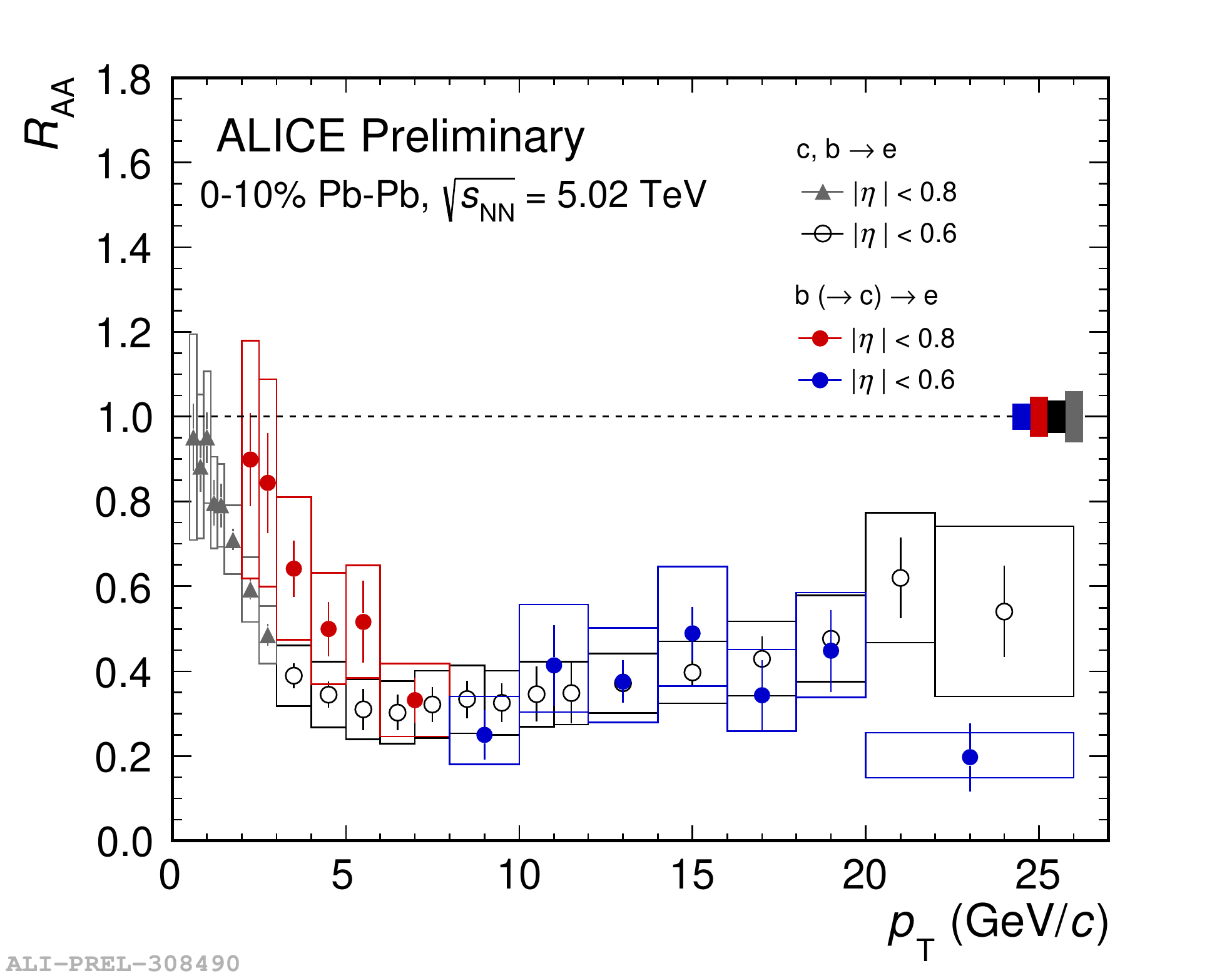}
\includegraphics[width=6cm]{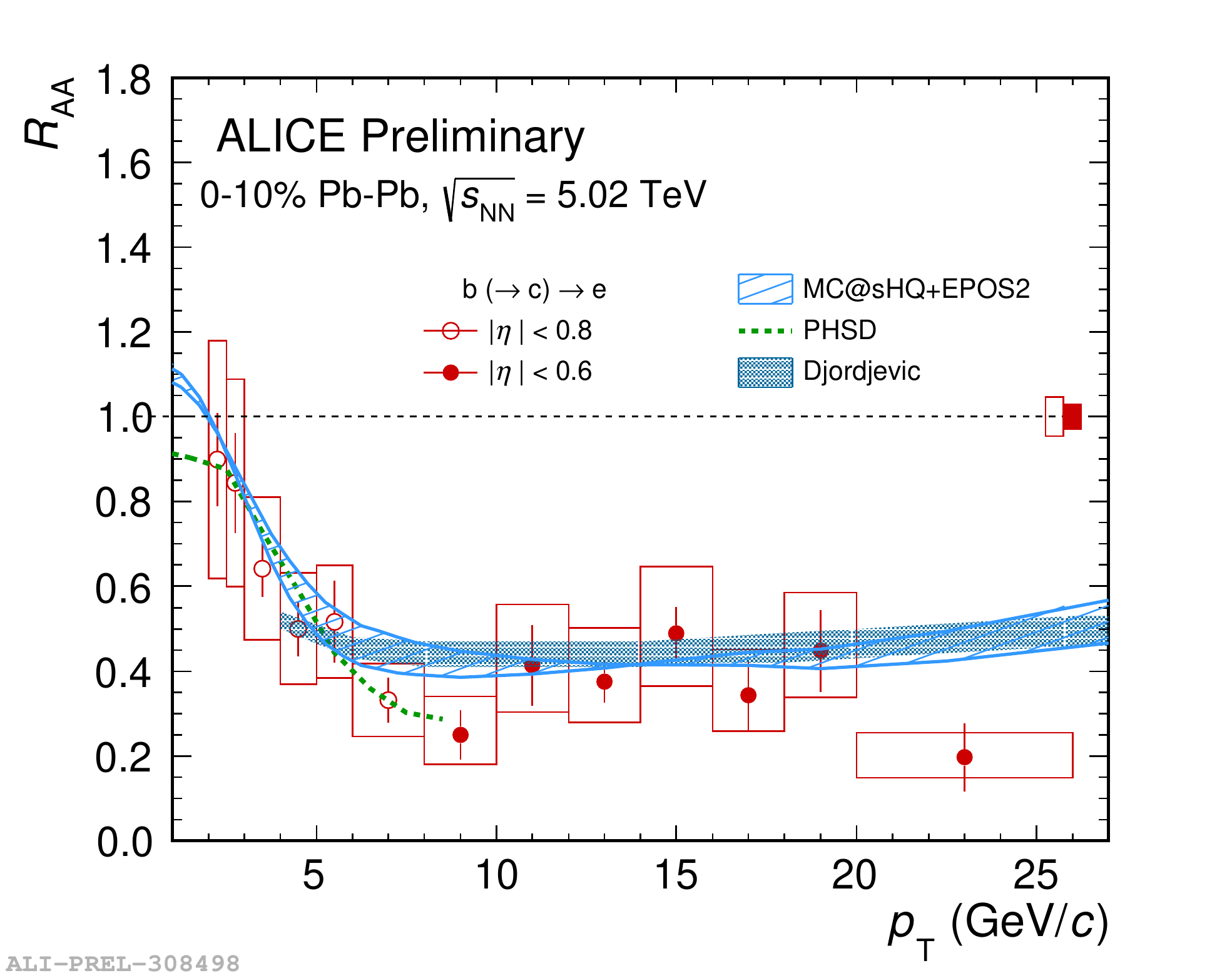}
\includegraphics[width=6cm]{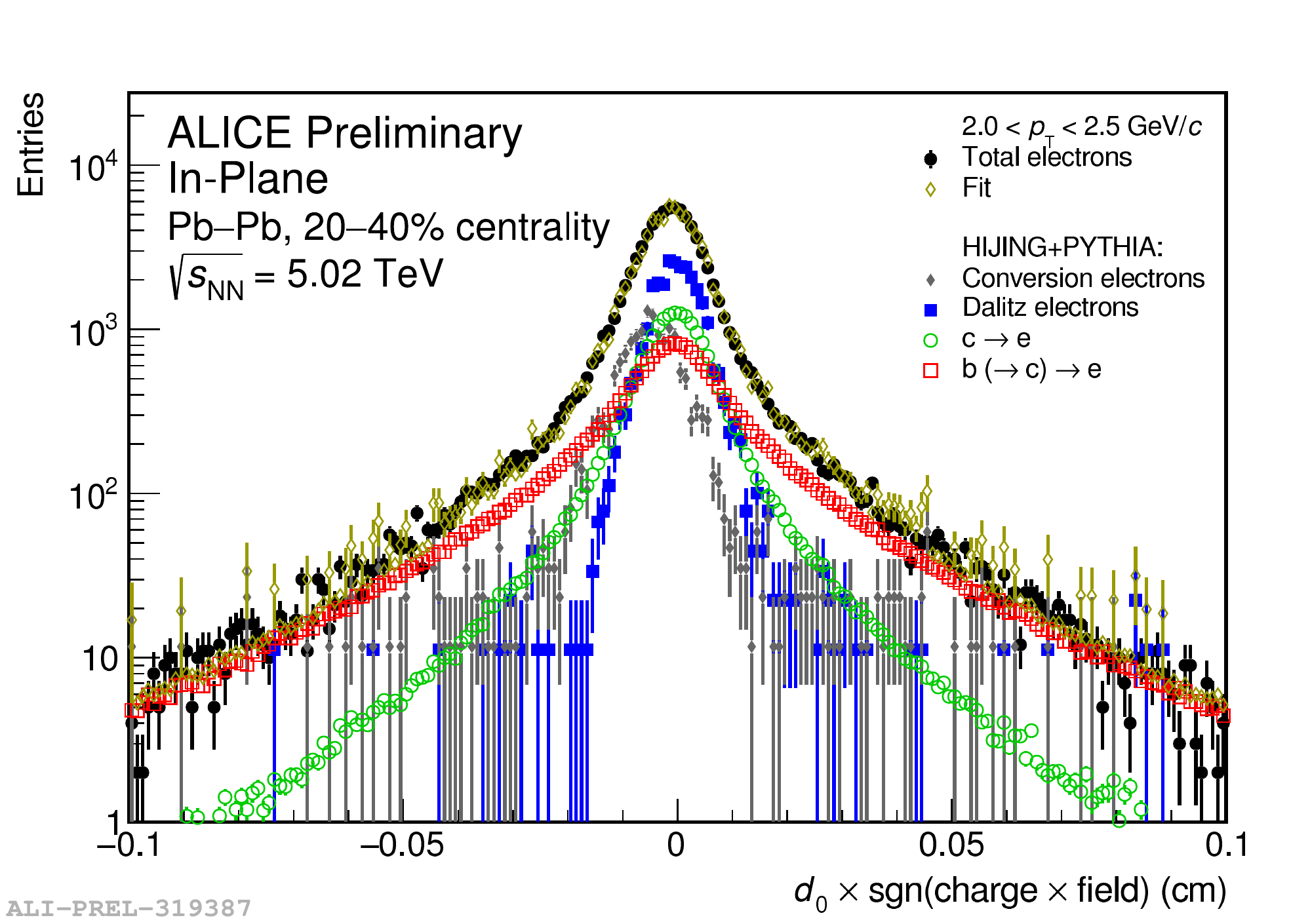}
\includegraphics[width=6cm]{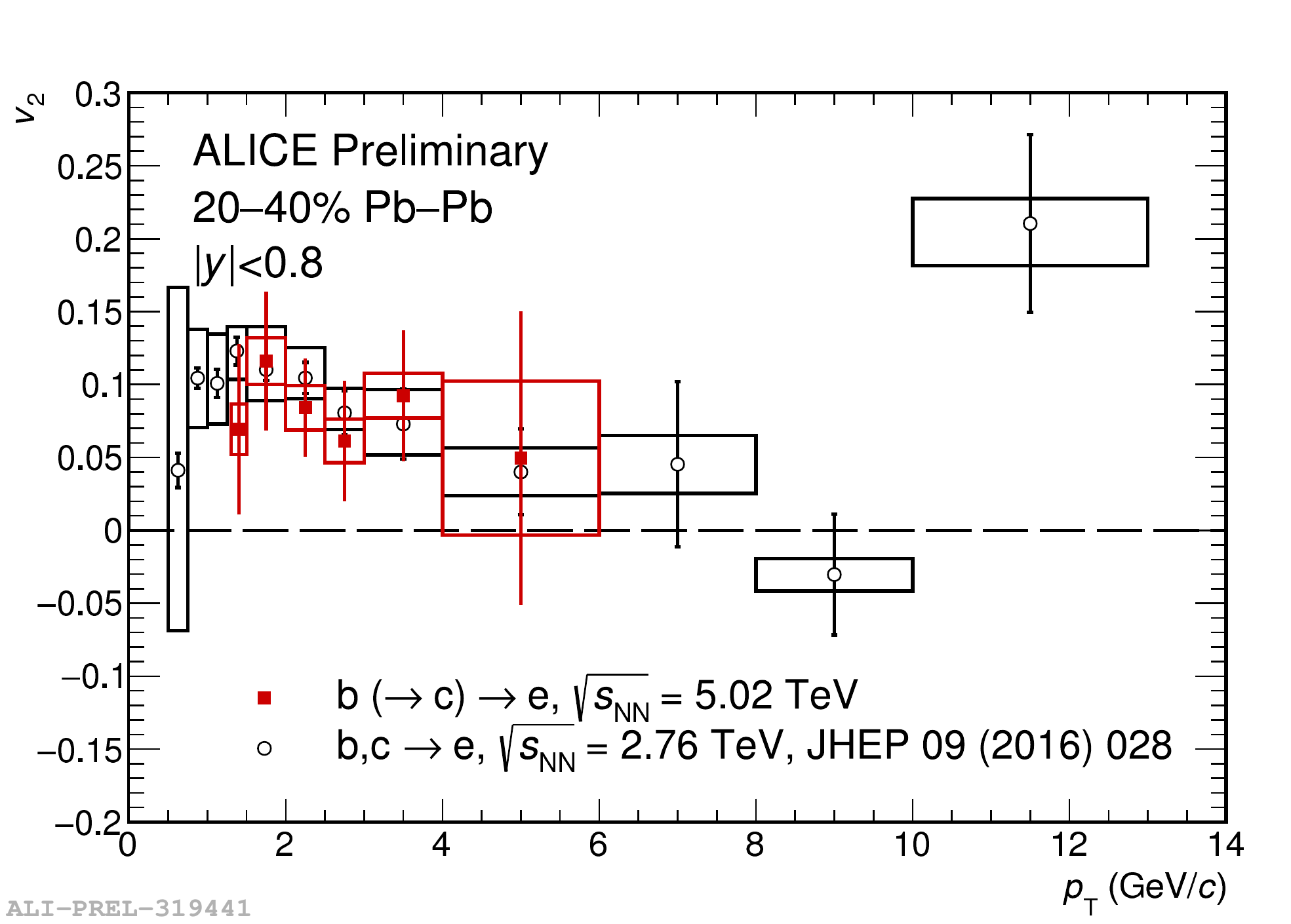}
\end{center}
\caption{\label{fig:be}Various results of beauty-hadron decay electrons in ALICE. The top two panels show the $R_{\mathrm{AA}}$ in 0--10\% Pb--Pb collisions. The bottom-left panel shows an example of a MC template fit of the $d_0$ of electrons to extract those from beauty-decay. The bottom-right panel shows the $v_{2}$ of beauty-hadron decay electrons in 20--40\% Pb--Pb collsions. }
\end{figure}

Electrons from beauty-hadron decays must be separated from other sources of electrons, such as photon conversion, Dalitz decays, and charm-hadron decays. This separation is achieved by exploiting the relatively long decay length of beauty hadrons ($\tau_{B} \approx$ 500 $\mu$m$/c$) versus non-beauty hadrons (e.g. \linebreak ${\tau_{D} = \text{60--300}\text{ }\mu \mathrm{m}/c}$) \cite{pdg}. The long decay length gives beauty-hadron decay electrons a longer distance of closest approach to the primary vertex ($d_{0}$), thus making the beauty-electron $d_{0}$ distribution much wider than that from other sources (Fig. \ref{fig:be}, bottom left). The difference in $d_{0}$ shape allows us to fit the electron $d_{0}$ distribution with templates from Monte Carlo simulations in order to extract the beauty-hadron decay electron yield. Once the yields are obtained in both pp and Pb--Pb collisions, the nuclear modification factor $R_{\mathrm{AA}} = \frac{\mathrm{d}N_{\rm{AA}}/\mathrm{d}p_{\rm{T}}}{T_{\rm{AA}}*\mathrm{d}\sigma_{\rm{pp}} /\mathrm{d}p_{\rm{T}}}$ can be calculated. The pp reference spectrum for the beauty-hadron decay electron $R_{\mathrm{AA}}$ measurement in Pb-Pb at $\sqrt{s_{\mathrm{NN}}} = 5.02$ TeV was obtained by a pQCD-driven scaling of the cross section measured at $\sqrt{s} = 7$ TeV \cite{bepp}. In order to obtain a $v_{2}$ measurement, the yield is measured for electrons that lie both in and out of the event plane, and the $v_{2}$ is calculated according to $v_{2} = \frac{1}{R^{2}} \frac{\pi}{4} \frac{N_{\mathrm{in-plane}}-N_{\mathrm{out-of-plane}}}{N_{\mathrm{in-plane}}+N_{\mathrm{out-of-plane}}}$, where $N$ refers to the number of electrons measured, and $R$ is the resolution correction for the event plane. 

Two beauty-electron measurements in ALICE are $R_{\mathrm{AA}}$ in 0--10\% and $v_{2}$ in 20--40\% central Pb--Pb collisions at $\sqrt{s_{\mathrm{NN}}} = 5.02$ TeV. In Fig. \ref{fig:be}, top left, the beauty-electron $R_{\mathrm{AA}}$ is shown along with the $R_{\mathrm{AA}}$ of heavy-flavor decay electrons (from both charm and beauty decays). Though the systematic error bars are large, a hint of an increased $R_{\mathrm{AA}}$ of beauty-hadron decay electrons when compared with heavy-flavor electrons at low-$p_\mathrm{T}$ is observed. This is consistent with our expectations of the mass dependence of energy loss in the QGP medium. At high $p_{\rm{T}}$, the two distributions overlap, in part because at higher momentum, the heavy-flavor electron sample becomes dominated by beauty-hadron decay electrons. Fig. \ref{fig:be} (top right) shows a comparison of the $R_{\mathrm{AA}}$ of beauty-hadron decay electrons with models that include both collisional and radiative energy loss. We see that the theoretical models are in good agreement with data. 

Finally, the $v_{2}$ of beauty-hadron decay electrons is shown in Fig. \ref{fig:be} (bottom right). The $v_{2}$ is non-zero; in fact, between $1.3 < p_{\mathrm{T}} < 4$ GeV/$c$, the significance of the measurement for a positive $v_2$ is 3.49$\sigma$. This hints that the beauty quark may participate in the collective behavior of the medium. The $v_{2}$ measurement of heavy-flavor decay electrons \cite{hfeV2}, also shown in Fig. \ref{fig:be}, is similar to that of electrons from beauty decays.
\subsection{Non-prompt D$^{0}$ mesons}
\begin{figure}[h]
\begin{center}
\includegraphics[width=6cm]{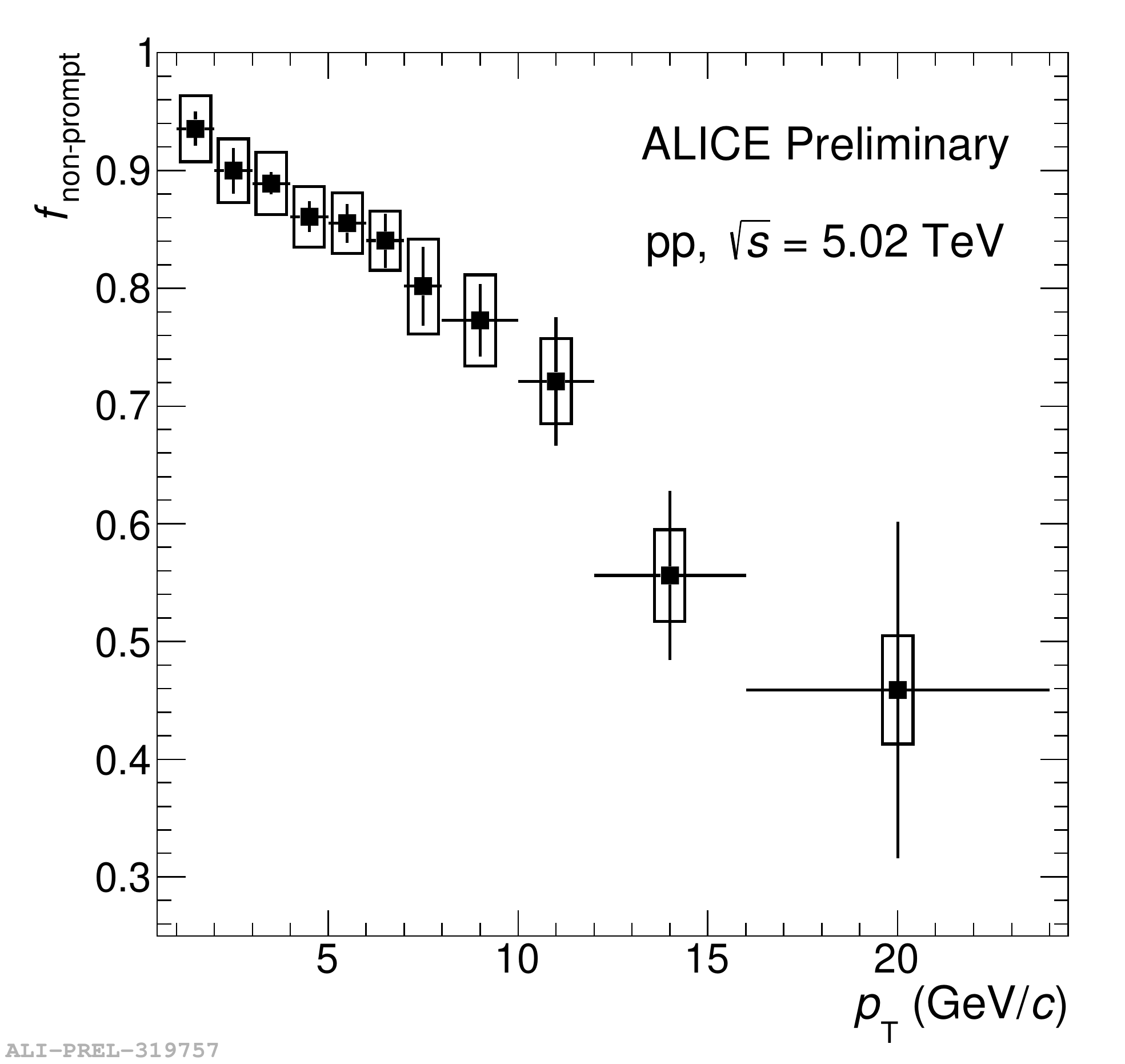}
\includegraphics[width=5.5cm]{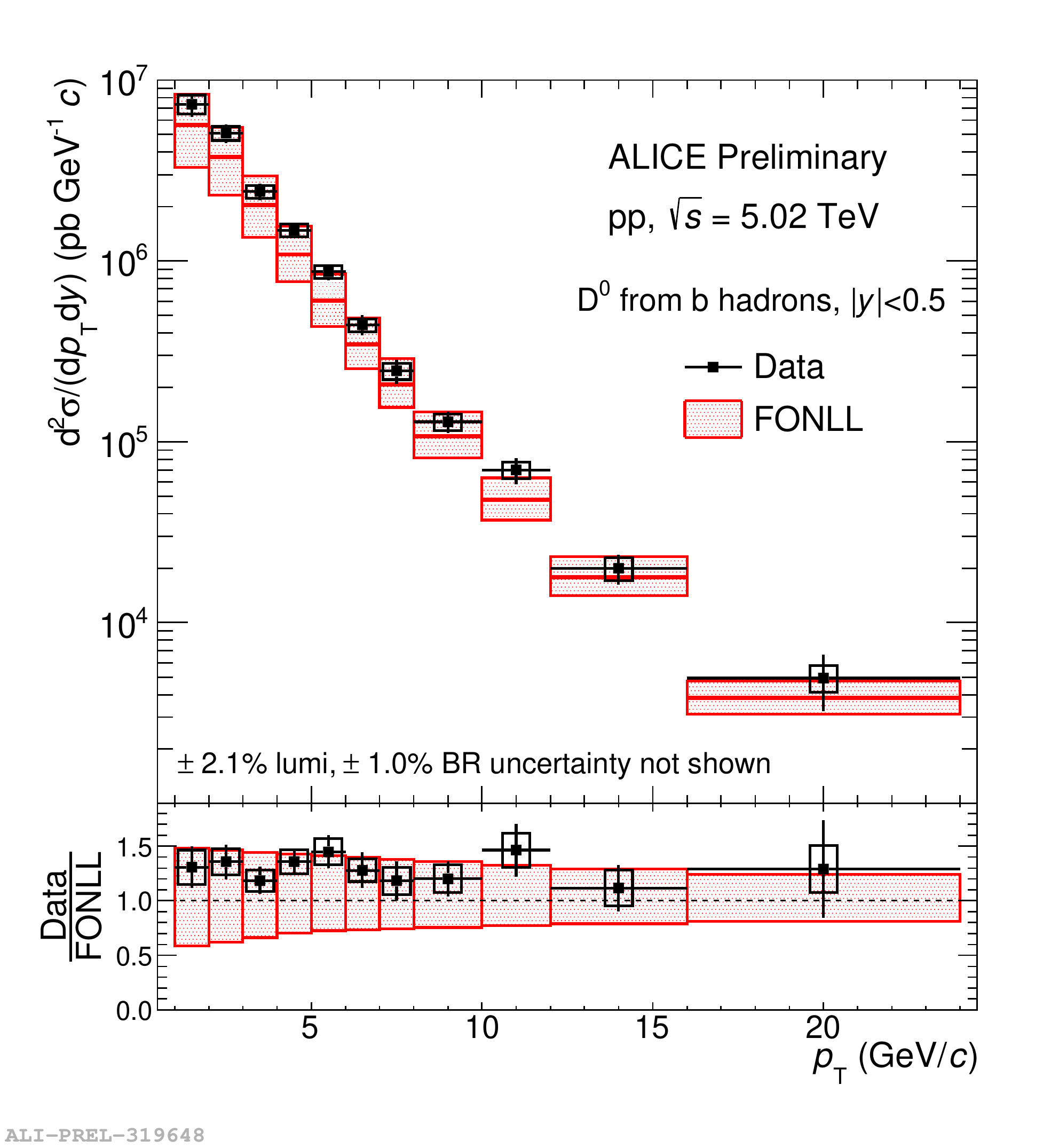}
\end{center}
\caption{\label{fig:nonPD0}Left: the fraction of non-prompt to inclusive D$^{0}$ mesons in the sample. Right: The cross section of non-prompt D$^{0}$ mesons in pp collisions at $\sqrt{s} =$ 5.02 TeV compared to FONLL \cite{fonll}.}
\end{figure}

With the ALICE detector, beauty production is also studied by measuring non-prompt D$^{0}$ mesons from beauty-hadron decays. The measurement is performed in pp collisions at $\sqrt{s} =$ 5.02 TeV. The non-prompt D$^{0}$ mesons (along with their charge conjugates) are reconstructed via the decay channel to K$^{-}\pi^{+}$ (branching ratio $\sim 3.9\%$) and selected by applying various topological requirements, including a cut on the distance between the primary and secondary vertices. The topological selection criteria are optimized using boosted decision trees, allowing us to achieve a high fraction of non-prompt D$^{0}$ mesons ($f_{\text{non-prompt}}$) in our sample. At low-$p_{\mathrm{T}}$, $f_{\text{non-prompt}}$ almost reaches 95\%, an unprecedented purity for this measurement (see Fig. \ref{fig:nonPD0}).
The cross-section of non-prompt D$^{0}$ mesons is compared with FONLL \cite{fonll} predictions (pQCD) in Fig. \ref{fig:nonPD0}. As in previous measurements \cite{cms}, the two are in agreement, though the measurement lies on the upper edge of the FONLL uncertainty band. 
\subsection{Beauty-tagged Jets}
A more direct access to the initial parton kinematics is obtained by measuring beauty-tagged jets. This has been done for the first time in ALICE in p--Pb collisions at $\sqrt{s_{\rm{NN}}} = 5.02$ TeV. Jets are selected using the anti-$k_{\mathrm{T}}$ algorithm \cite{antikt}, and a resolution parameter of $R=$ 0.4.  To achieve a high purity of b-jets in the sample, the long lifetime of beauty hadrons is exploited once more. Jets that contain a three-pronged secondary vertex are selected, and a number of topological requirements are applied to increase the b-jet purity. In particular, a cut is applied on the displacement significance ($SL_{xy} >$ 7) of the secondary vertex. The $SL_{xy}$ is defined as the distance between primary and secondary vertex in the $xy$-plane divided by the resolution of that distance. 
\begin{figure}[h]
\includegraphics[width=6cm]{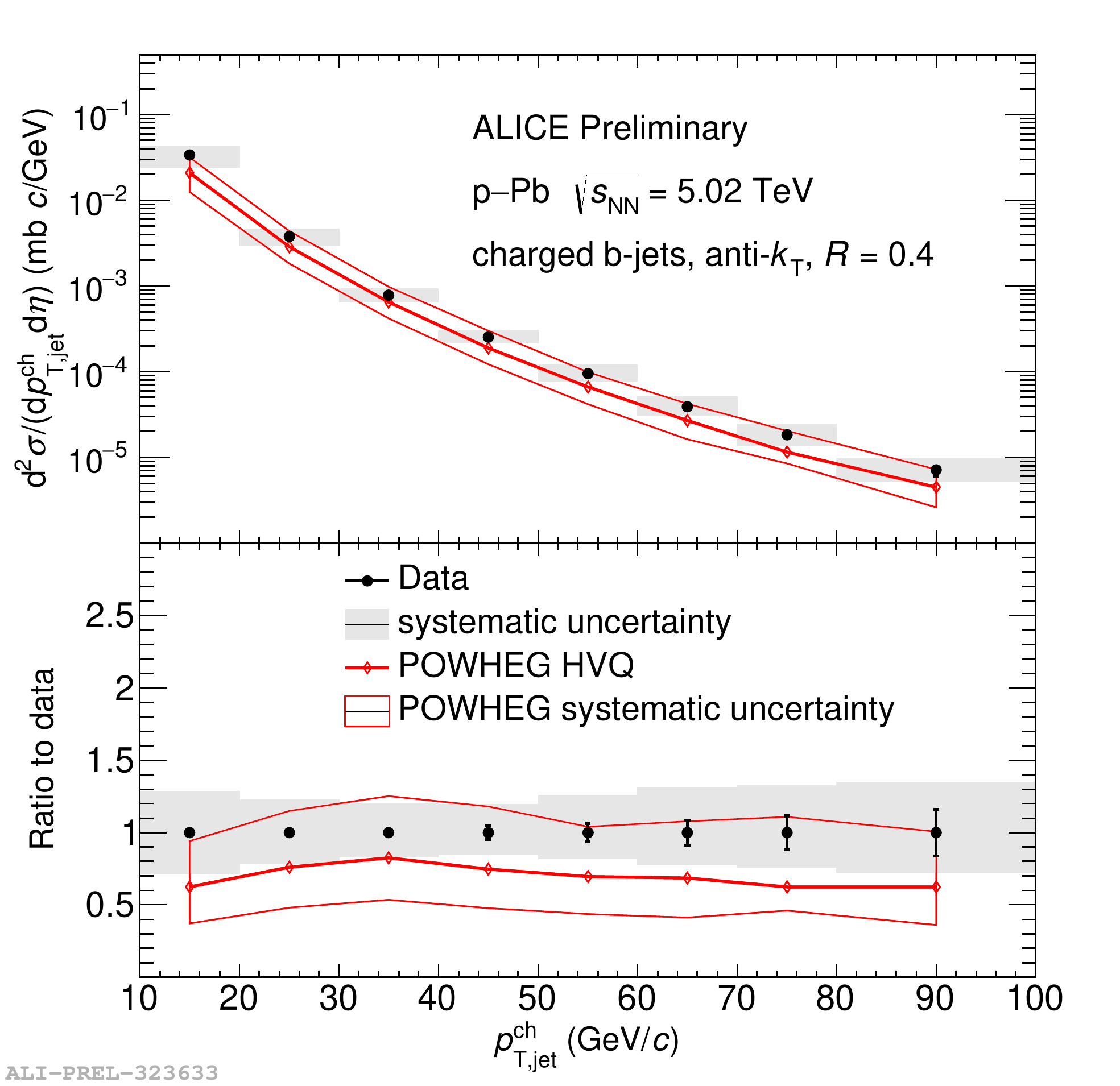}
\includegraphics[width=6cm]{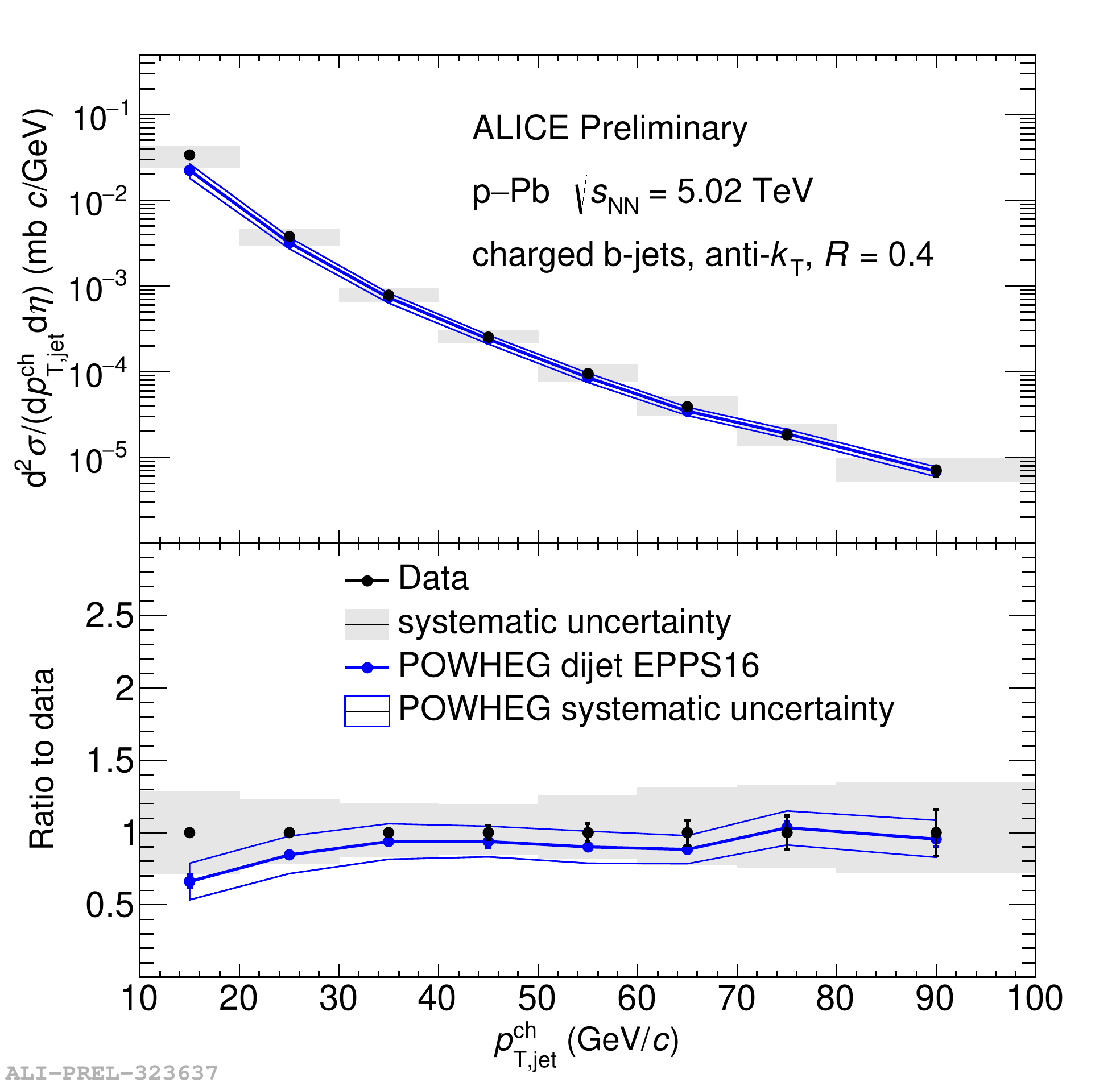}
\begin{center}
    \includegraphics[width=8cm]{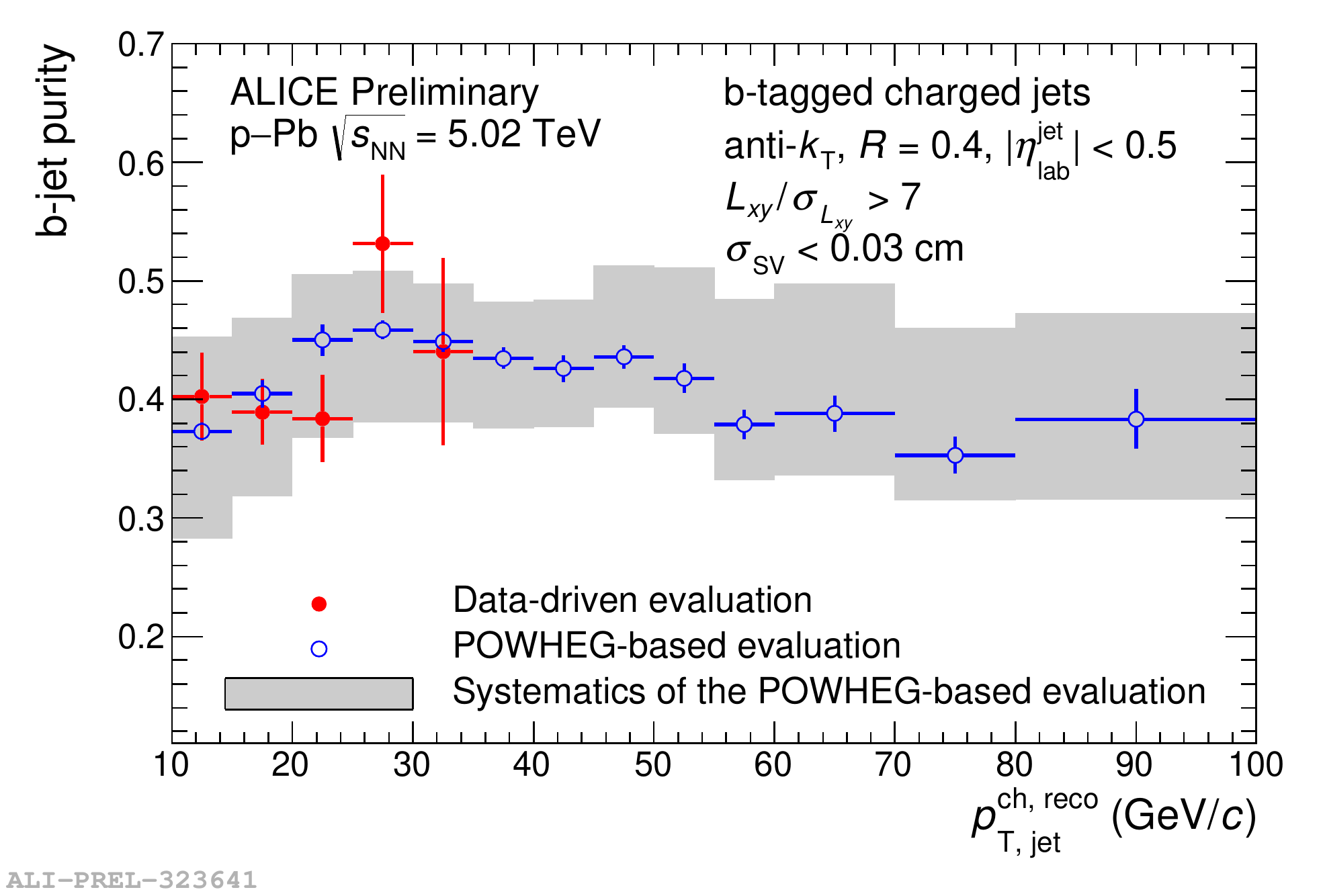}
\end{center}
\caption{\label{fig:bJets}Top left and right: the cross-section of beauty-tagged jets in p--Pb collisions compared with POWHEG. Bottom: the purity of beauty jets in the sample.}
\end{figure}
The results of this measurement are shown in Fig. \ref{fig:bJets}, where the estimated purity ($\sim$40\%) of beauty-tagged jets in the sample is shown along with the jet cross-section compared POWHEG HVQ and POWHEG dijet EPPS16. We see that both models agree with the measured cross-section. 
\section{Conclusions}
In this manuscript the ALICE results on beauty production were discussed with particular focus on beauty-hadron decay electrons in Pb--Pb collisions, non-prompt D$^{0}$ mesons in pp collisions, and beauty-tagged jets in p--Pb collisions. All three analyses took advantage of the long lifetime of beauty hadrons to separate the beauty signal from background sources. The non-prompt D$^{0}$ meson measurement agrees with FONLL predictions, and the beauty-tagged jet measurements agree with POWHEG models. In Pb--Pb collisions, we see a hint of the mass-dependent energy loss in the QGP, as well as a non-zero $v_{2}$ for beauty-hadron decay electrons. 

\paragraph{Acknowledgements:}
This work was supported by U.S. Department of Energy Office of Science under contract number
DE--SC0013391.

%

\end{document}